\documentclass[11pt]{article}%
\usepackage{amsmath}
\usepackage{graphicx}
\usepackage{cite}
\usepackage{subfigure}
\usepackage{multirow}
\usepackage{authblk}
\usepackage{indentfirst}
\usepackage{caption}
\usepackage{algorithm}
\usepackage{algorithmicx}
\usepackage{algpseudocode}
\usepackage{graphicx}
\usepackage{bm}
\usepackage{CJK}
\usepackage{geometry}
\usepackage{float}
\usepackage{diagbox}
\usepackage{colortbl}
\usepackage{textcomp,booktabs}
\usepackage{array}
\usepackage{xcolor}%
\usepackage{amsfonts}%
\usepackage{amssymb}
\newcommand{\eqnb}{\begin{equation}}
\newcommand{\eqne}{\end{equation}}

\newtheorem{The}{Theorem}

\newtheorem{Lem}{Lemma}

\newtheorem{Rem}{Remark}
\definecolor{mygray}{gray}{.9}
\definecolor{mypink}{rgb}{.99,.91,.95}
\definecolor{mycyan}{cmyk}{.3,0,0,0}

\begin{document}
\title{Information Theory of Blockchain Systems}
\author{Quan-Lin Li$^{a}$, Yaqian Ma$^{a}$, Jing-Yu Ma$^{b*}$, Yan-Xia Chang$^{a}$\\$^{a}$School of Economics and Management,\\Beijing University of Technology, Beijing 100124, China\\$^{b}$Business School, Xuzhou University of Technology, Xuzhou 221018, China}
\maketitle

\begin{abstract}
In this paper, we apply the information theory to provide an approximate expression of the steady-state probability distribution for blockchain systems. We achieve this goal by maximizing an entropy function subject to specific constraints. These constraints are based on some prior information, including the average numbers of transactions in the block and the transaction pool, respectively. Furthermore, we use some numerical experiments to analyze how the key factors in this approximate expression depend on the crucial parameters of the blockchain system. As a result, this approximate expression has important theoretical significance in promoting practical applications of blockchain technology. At the same time, not only do the method and results given in this paper provide a new line in the study of blockchain queueing systems, but they also provide the theoretical basis and technical support for how to apply the information theory to the investigation of blockchain queueing networks and stochastic models more broadly.

\textbf{Keywords:} Blockchain; Information theory; Maximum entropy principle; Steady-state probability distribution. 
\end{abstract}

\section{Introduction}

Blockchain has become a prominent topic of discussion in recent years, revolutionizing various aspects of life through its significant impact on many practical application fields. For example, finance by Kowalski et al. \cite{Tsa:2016}; the Internet of Things by Torky and Hassanein \cite{Huc:2016}; healthcare by Sudeep et al. \cite {Met:2016}; and others. The active participation of miners in the mining process is fundamental to ensuring the secure and stable operation of the blockchain system, as well as guaranteeing its sustainable development. However, the inner workings of blockchain mining are extremely obscure and challenging to examine. Conducting direct measurements on mining networks is highly complex due to the miners' privacy concerns, whereas blockchain data provides a method of direct measurement. Consequently, it is essential to develop statistical techniques using accessible blockchain data for investigating blockchain systems.

So far blockchain research has obtained many important advances, readers may refer to
a book by Swan \cite{Swa:2015}; a key research framework shown by Daneshgar et al. \cite{Dan:2019}, Lindman et al. \cite{Lind:2017} and Risius and Spohrer \cite{Ris:2017}; decision in blockchain mining by Ma and Li \cite{Cat：2016} and Chen et al. \cite{Chr：2020};
and others by Lu et al. \cite{Lu:2020} and Yang et al. \cite{Yan:2020}. 

Applying queueing theory and Markov processes to analyze blockchain systems is interesting but challenging, since each blockchain system not only is a complicated stochastic system but also has multiple key factors and
a physical structure with different levels. Li et al. \cite{Li:2018} provided a two-stage queueing model of the PoW blockchain system, clearly described and expressed the physical structure with multiple key factors, furthermore the matrix geometric solution was applied to give a complete solution such that the performance evaluation of the PoW blockchain system was established in a simple form. Seol et al. \cite{ref14} proposed an $M(1,n)/M_{n}/1$ queueing model to analyze the blockchain system in Ethereum; Zhao et al. \cite{ref15} established a non-exhaustive queueing model with a limited
batch service and a possible zero-transaction service, derived the average number of transactions and the average confirmation time of a transaction; Mi$\breve{s}$i$\acute{c}$ et al. \cite{ref16} applied the Jackson network to analyze the blockchain network. 

Compared with the queueing theory, the Markov process is mainly used to evaluate the throughput, confirmation time, security and privacy protection of the blockchain systems. Huang et al. \cite{Hua:2019} proposed the Markov process with an absorption state and conducted an analysis on the performance of the Raft consensus algorithm in private blockchains. Srivastava \cite{Sri:2019} calculated the transaction confirmation time in blockchain systems. Li et al. \cite{Li:2021} discussed block access control mechanisms in wireless blockchain networks. Nguyen et al. \cite{Ngu:2020} investigated the task offloading problem in mobile blockchain with privacy protection using Markov processes and deep reinforcement learning.

The traditional reluctance of miners to share insider information regarding their competitive advantages, leading to great difficulties for these two approaches when dealing with more complex blockchain systems, such as those involving multiple mining pools. The purpose of this paper is to apply the maximum entropy principle to provide an approximate expression for blockchain systems. In information theory, entropy serves as a probabilistic measure to quantify the uncertainty of information associated with random variables. In recent years, the information entropy has been implemented in various practical domains of blockchain technology. For example, industrial Internet of Things by Khan and Byun \cite{Khan:2020}; renewable energy by Liu et al. \cite{Liu:2023}; fake news prevention by Chen et al. \cite{Chen:2022}; and medical data sharing by Liang et al. \cite{Lia:2019}. 

The degree of randomness in a random variable can be measured by applying maximum entropy when its information is most uncertain. For example, a large amount of information can only be partially obtained and utilized. For random variables, Jaynes \cite{Jay:1957a, Jay:1957b} initially proposed the maximum entropy principle, which offers an approximate computational approach for unknown probability distributions. Such an approach provides a uniquely correct self-consistent method of inference for estimating probability distributions based on the available information. 

The main contributions of this paper are twofold. The first one is to apply the maximum entropy principle to study blockchain queueing systems for the first time. Different from previous works for applying queueing theory or Markov processes, we just need to take statistical techniques by simple observation on miners. The second contribution of this paper is to provide the approximate expression of the steady-state probability distribution for blockchain systems. So far, numerous categories of blockchain systems have yet to be thoroughly analyzed using queueing theory or Markov processes due to difficulties in the expression of the steady-state probability distributions. Therefore, the results of this paper give new insights into applying the maximum entropy principle to more complex blockchain systems. For example, the PoW blockchain system with multiple mining pools, the PBFT blockchain system of dynamic nodes, the DAG-based blockchain systems, the Ethereum, and the large-scale blockchain systems with either cross-chain, side-chain, or off-chain. 

The rest of this paper is organized as follows. Section 2 introduces the blockchain queueing model briefly. In Section 3, we apply the maximum entropy principle to give the approximate expression of the steady-state probability distribution for the blockchain system. We also conduct numerical experiments to analyze how the key factors of the approximate expression depend on some crucial parameters in Section 4. Finally, the whole work is concluded in the last section.
 
\section{Model Distribution}
In this section, we describe a blockchain system as two stages of asynchronous processes: block-generation and blockchain-building, which is depicted in Fig. 1. To ensure clarity, we review the blockchain queueing model and adopt the notations of Li et al. \cite{Li:2018} briefly . 
\begin{figure}[htbp]
\centering        \includegraphics[width=12.5cm]{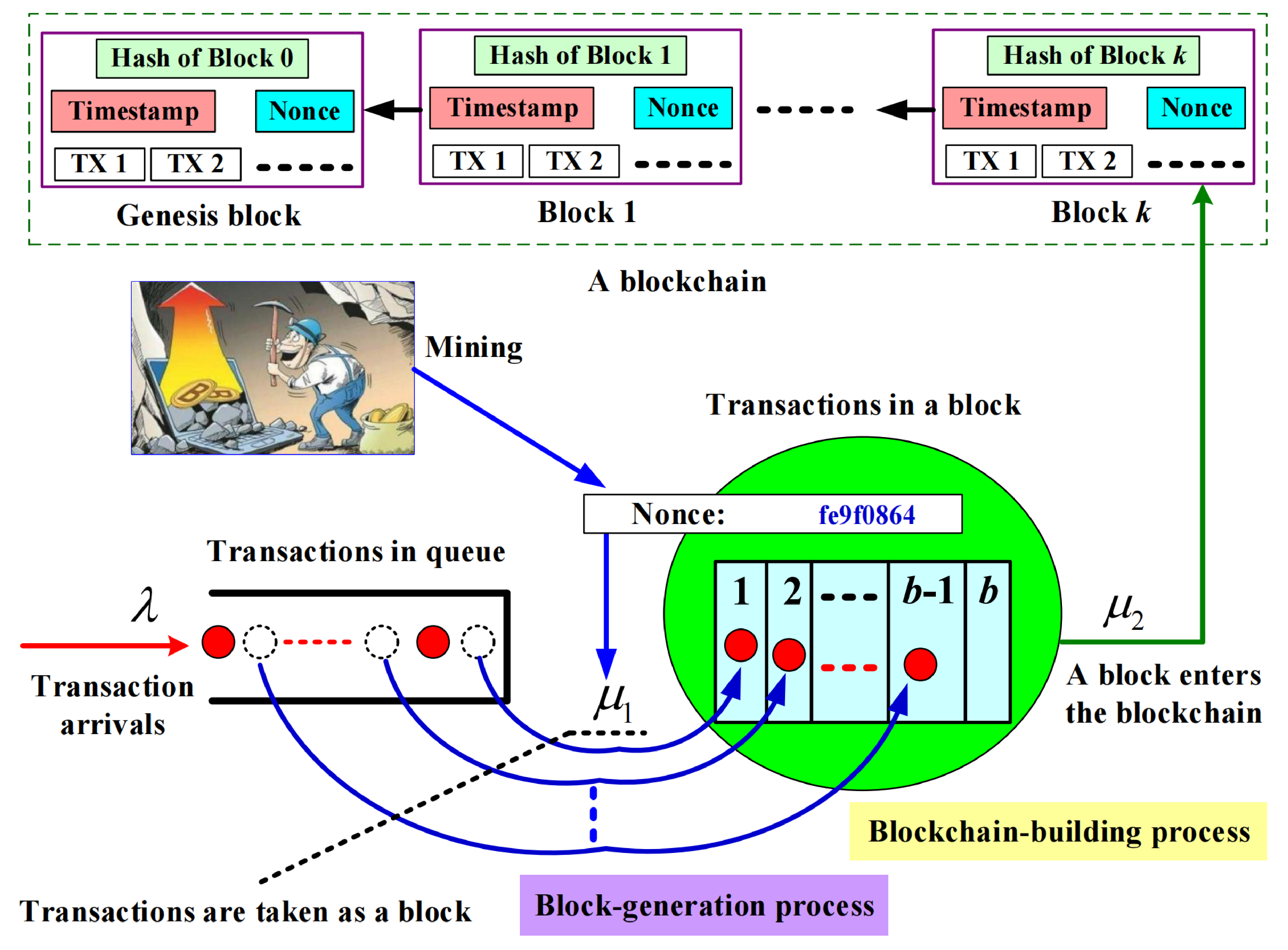}
\caption{A blockchain queueing system.}%
\end{figure}

\textbf{Arrival processes}: Transactions arrive at the blockchain system according to a Poisson process with arrival rate $\lambda$. Each transaction must first enter and queue up in a transaction pool with infinite size. 

\textbf{Block-generation processes}: Each arrival transaction first queues up in the transaction pool and then waits to be mined into a block successfully.  We assume that the block-generation times are i.i.d. and exponential with service rate $\mu_{1}$. The transactions are chosen into the block, but they are not completely based on the First Come First Service (FCFS) from the order 
of transaction arrivals.

\textbf{Block capacity}: To avoid the spam attacks, we assume that the maximum size of each block is limited to $b$ transactions. If there are more than $b$ transactions in the transaction pool, then the $b$ transactions are selected to form a full block while the rest of transactions are still waiting in the transaction pool and may be used to construct another block. 

\textbf{Blockchain-building processes}: The block with a group of transactions will be pegged to a blockchain. We assume that the blockchain-building times are i.i.d. and exponential with the service rate $\mu_{2}$. 

\textbf{Independence}: We assume that all the random variables defined above are independent of each other.

Let $I(t)$ and $J(t)$ be the numbers of transactions in the block and in the transaction pool at time $t$, respectively. Then, $(I(t),J(t))$ may be regarded as a state of the blockchain system at time $t$. The state space of this blockchain system is
\[ \mathbf{\Omega}  =   \left\{ \left( i, j  \right) , 0 \leq i \leq b , 0 \leq j \leq \infty \right\}. \]

%Let $X\left(  t\right)  =\left(  I\left(  t\right)  ,J\left(  t\right)
%\right)  $. Then $\left\{  X\left(  %t\right)  :t\geq0\right\}  $ is a
%continuous-time Markov process on the state space $\mathbf{\Omega}$.

The following lemma provides a necessary and sufficient condition under which the blockchain system is stable. Here, we only restate it without proof, while readers may refer to  Chapter 3 of Li \cite{Li:2010} and Li et al. \cite{Li:2018} for more details.
\begin{Lem}
The blockchain system is stable if and
only if
\begin{equation} \label{1}
	\dfrac{b \mu_{1} \mu_{2}}{\mu_{1} + \mu_{2}}  > \lambda.
\end{equation}
\end{Lem}

In what follows we assume that the stable condition (\ref{1}) is satisfied, then this blockchain system is stable. The limit \[\lim_{t \to +\infty} p \left \lbrace I(t)=i,J(t)=j \right\rbrace \] exists and is unique. Let
\[p(i,j)=\lim_{t \to +\infty} p \left\lbrace I(t)=i,J(t)=j \right\rbrace .\]
Then, $p(i,j), (i,j) \in \mathbf{\Omega}$ is the steady-state probability distribution of the blockchain system.

By using the matrix-geometric solution, we can write the steady-state probability distribution under the stable condition (\ref{1}), see Li et al. \cite{Li:2018}. In the next section, we will introduce the maximum entropy principle to provide the approximate expression of the steady-state probability distribution for the blockchain system.

\section{Maximum Entropy in Blockchain Systems}
In this section, we provide an entropy function and some prior information, and use Lagrange method of undetermined multipliers to give the approximate expression of the steady-state probability distribution.
\subsection{Entropy Function}
Based on the steady-state probability distribution $p(i,j)$, we introduce the entropy function 
\[ H(p) = - \sum_{(i,j) \in \mathbf{\Omega}} p(i,j) \ln p(i,j) \]
or
\begin{equation} \label{2}
	H(p) = - \sum_{j=0}^{\infty} \sum_{i=0}^{b} p(i,j) \ln p(i,j).
\end{equation}

The maximum entropy principle states that of all distributions satisfying the constraints supplied by the given information, the minimally prejudiced distribution $p(i,j),(i,j) \in \mathbf{\Omega}$ is the one that maximizes the entropy function of the blockchain queueing system.
\subsection{Prior information}
To approximate the steady-state probability distribution $p(i,j),(i,j) \in \mathbf{\Omega}$ using the maximum entropy principle by maximizing (\ref{2}), we need to provide some prior information as follows:

(i) The normalisation:
\begin{equation}
\sum\limits_{(i,j) \in \mathbf{\Omega} } p(i,j) = 1.\label{3}
\end{equation}

(ii) The average number of transactions in the block:
\begin{equation}
\sum\limits_{i=0}^{b} i \sum_{j=0}^{\infty} p(i,j) = I.\label{4}
\end{equation}

(iii) The average number of transactions in the transaction pool:
\begin{equation}
\sum\limits_{j=0}^{\infty} j \sum_{i=0}^{b} p(i,j) =J.\label{5}
\end{equation}

\begin{Rem}
Note that the statistics of prior information selected always may be known numerically via system measurements during finite observation periods or can be determined symbolically via known analytic formulae based on operational or stochastic assumptions. For example, blockchain data has the advantage of providing direct measurements, as the fields of a block are filled by the miner of that block.
\end{Rem}

\subsection{The maximum entropy principle}
The steady-state probability distribution $p(i,j)$ is considered as an independent variable. We maximize the entropy function (\ref{2}) subject to constrains (\ref{3})-(\ref{5}), the optimization model of the maximum entropy principle can be written as 
\begin{align*}
&\max\,\, H(p) = - \sum_{j=0}^{\infty} \sum_{i=0}^{b} p(i,j) \ln p(i,j), \\
&s.t.\quad
\begin{cases}
\sum\limits_{j=0}^{\infty}\sum\limits_{i=0}^{b}p(i,j) = 1, \\
\sum\limits_{i=0}^{b} i \sum\limits_{j=0}^{\infty} p(i,j) = I,\\
\sum\limits_{j=0}^{\infty} j \sum\limits_{i=0}^{b} p(i,j) =J. \\
\end{cases}
\end{align*}

The following theorem provides the approximate expression of the steady-state probability distribution for the blockchain system by the maximum entropy principle. 
\begin{The}
    \noindent  For the steady-state probability distribution  $p(i,j)$ of blockchain systems, there exists a tuple of positive numbers $x$, $y$ and $z$ that satisfy
	\[ \tilde{p}(i,j) = xy^{i}z^{j}.\]
\end{The}

Proof: By introducing $\beta_{0}$, $\beta_{1}$ and $\beta_{2}$ to equations (\ref{3})-(\ref{5}), we write Lagrangian function as
\begin{align}
		L(p,\beta_{0},\beta_{1},\beta_{2})	=&  - \sum_{j=0}^{\infty} \sum_{i=0}^{b} p(i,j) \ln p(i,j)  		 + \beta_{0} \left( 1-\sum_{j=0}^{\infty} \sum_{i=0}^{b} p(i,j) \right) \nonumber\\ 
		& + \beta_{1} \left( I-\sum\limits_{i=0}^{b} i \sum_{j=0}^{\infty} p(i,j)\right) 
		+ \beta_{2} \left(J-\sum\limits_{j=0}^{\infty} j \sum_{i=0}^{b} p(i,j)  \right),\label{6}
\end{align}
where $\beta_{0}$, $\beta_{1}$ and $\beta_{2}$ are the Lagrange multipliers corresponding to constraints (\ref{3})-(\ref{5}), respctively.

To find the maximum entropy solution $p(i,j)$, maximizing (\ref{2}) subject to constraints (\ref{3})-(\ref{5}) is equivalent to maximizing (\ref{6}).

The Lagrangian function $L(p,\beta_{0},\beta_{1},\beta_{2})$ is a multivariate function with respect to variables $p(i,j)$, $\beta_{0}$, $\beta_{1}$ and $\beta_{2}$. To obtain the maximum entropy solutions, we take the partial derivatives of $L(p,\beta_{0},\beta_{1},\beta_{2})$ with respect to $p(i,j)$ and then set the results equal to zero, i.e., $\partial L/ \partial p(i,j) = 0$. 

If $(i,j)$ is determined, then \[  \frac{\partial}{\partial p(i,j)} \left[ - \sum_{j=0}^{\infty} \sum_{i=0}^{b} p(i,j) \ln p(i,j)\right] = -\ln p(i,j) - 1.\]
It is clear that for all $(\overline{i},\overline{j})$, $\overline{i} \neq i $ and $\overline{j} \neq j$, 
\[ \frac{\partial}{\partial p(i,j)}   p(\overline{i},\overline{j}) \ln p(\overline{i},\overline{j}) = 0. \]
Thus, we obtain \[ \frac{\partial L}{\partial p(i,j)} = \left[ -\ln p(i,j) - 1\right]  - 
 \beta_{0} -   \beta_{1} i   -  \beta_{2} j =0, \]
which indicates
\begin{equation} \label{7}
\ln p(i,j) = -1 -\beta_{0} - \beta_{1} i -\beta_{2} j.
\end{equation} 
It follows from (\ref{7}) that 
\begin{equation} \label{7-1}
p(i,j) = \exp \left[ -\left( 1 +\beta_{0}\right) \right]\exp \left( - \beta_{1} i \right)\exp \left(-\beta_{2} j\right). 
\end{equation} 

Let 
\[ x =\exp \left[ -\left( 1 +\beta_{0}\right) \right], y=\exp \left( - \beta_{1}  \right) \text{ }\text{and}\text{ }  z=\exp \left(-\beta_{2} \right).\]
%where $x$ is the normalising constant.
Then, we rewrite (\ref{7-1}) as
\begin{equation} \label{8}
p(i,j) = x y^i z^j. 
\end{equation}
Substituting (\ref{8}) into (\ref{3}) and utilizing algebraic knowledge, we have
\begin{equation} \label{10}
 x = \frac{(1-y)(1-z)}{1-y^{b+1}}.
\end{equation}
Similarly, substituting (\ref{8}) into (\ref{4}) and (\ref{5}), respectively, we have
\begin{equation} \label{12}
 y^{b+1} -\sum_{n=1}^{b} \frac{1}{b-I}y^{n}+\frac{I}{b-I}=0   
\end{equation}
and 
\begin{equation} \label{11}
z=\frac{J}{1+J}.
\end{equation}

Therefore, if the average number of transactions in the block and the transaction pool can be provided, respectively, the positive numbers $x$, $y$ and $z$ exist to give the approximate 
expression for $\tilde{p}(i,j)$. This completes the proof.

\begin{Rem}
The theoretical expressions of the mean values $I$ and $J$ given by Li et al.  \cite{Li:2018} are restricted to Poisson arrival processes and exponential service times, meaning that these expressions are only theoretically applicable in this particular case. Nevertheless, the maximum entropy principle is not dependent on this assumption of the Poisson arrival processes and the exponential service times. It can be applied to non-Poisson arrival processes and non-exponential service times, as long as $I$ and $J$ can be provided, the non-linear equations can be solved to derive the approximate expression of the steady-state probability distribution for the blockchain queueing system. Therefore, the approximate expression derived in Section 3.3 has broad applicability. 
\end{Rem} 

\section{Numerical experiments}
In this section, we provide some numerical examples to verify computability of our theoretical results and analyze how the key factors $y$ and $z$ of the approximate expression depend on some crucial parameters of the blockchain queueing system.

Taking the situation of the Poisson arrival processes and the exponential service times in Li et al. \cite{Li:2018} as an example, since the theoretical expressions of the mean values $I$ and $J$ are composed of the crucial parameters $\lambda$, $\mu_{1}$, $\mu_{2}$ and $b$, we can observe the relation between the key factors and crucial parameters. Note that $x$ is represented by $y$ and $z$ according to equations (\ref{10})-(\ref{11}), we just need to focus on how $y$ and $z$ depend on these crucial parameters through numerical examples. 

In the Examples 1 and 2, we take some common parameters: The maximum block size $b=80$, the block-generation service rate $\mu_{1}=6, 7.5, 10$, blockchain-building service rate $\mu_{2}=2$ and the arrival rate $\lambda\in\left(  1, 3.5\right)$.

\textbf{Example 1} We analyze how $y$ depends on $\lambda$ and $\mu_{1}$. From Fig. 2, it is seen that $y$ decreases as $\lambda$ increases, while it also decreases as $\mu_{1}$ increases. 
\begin{figure}[htbp]
\centering        \includegraphics[width=9cm]{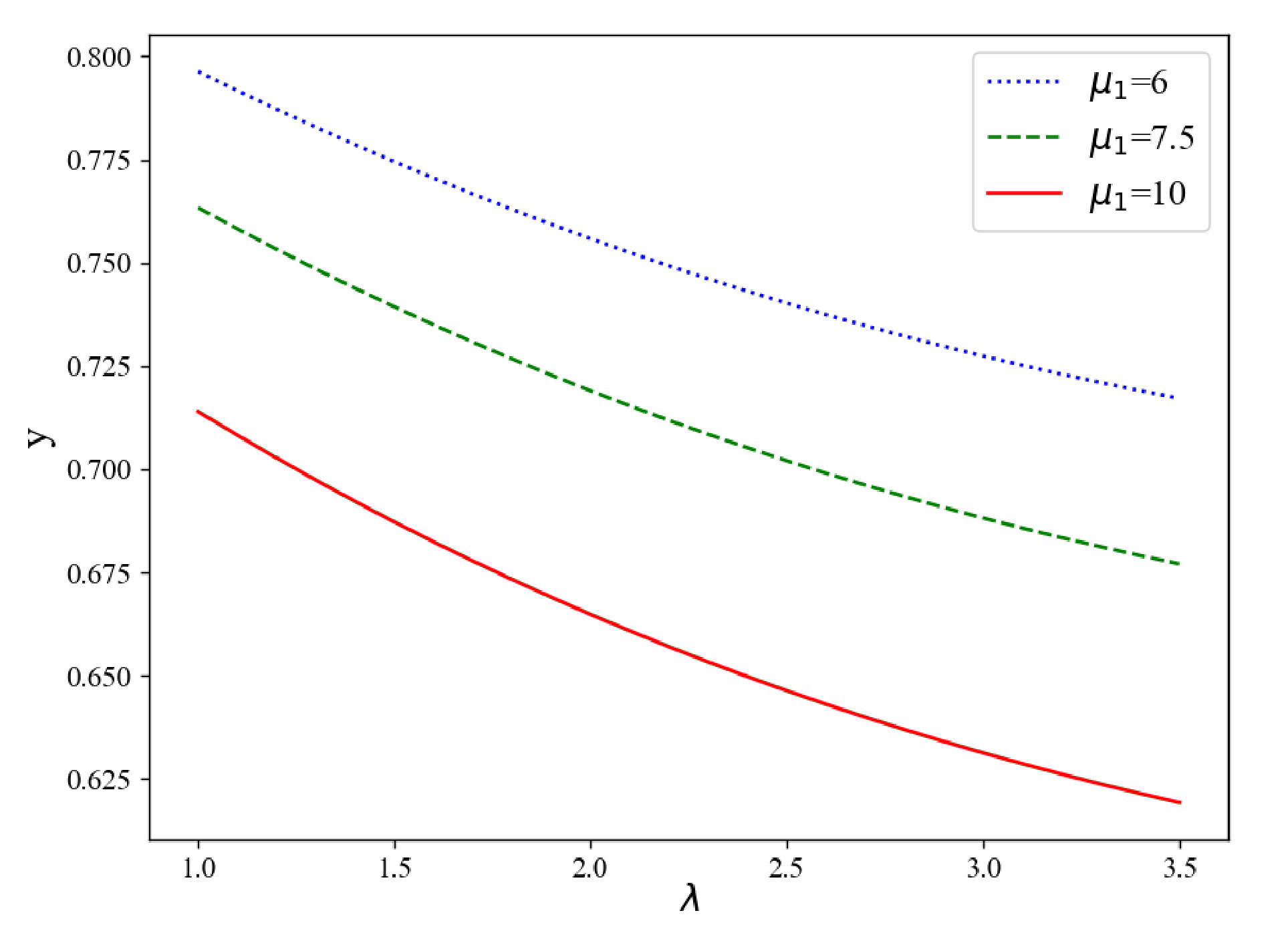}
\caption{$y$ vs. $\lambda$ for three different values of $\mu_{1}$.}%
\end{figure}

\textbf{Example 2} We analyze how $z$ depends on $\lambda$ and $\mu_{1}$. From Fig. 3, it is seen that $z$ increases as $\lambda$ increases, while it increases as $\mu_{1}$ decreases. 
\begin{figure}[htbp]
\centering        \includegraphics[width=9cm]{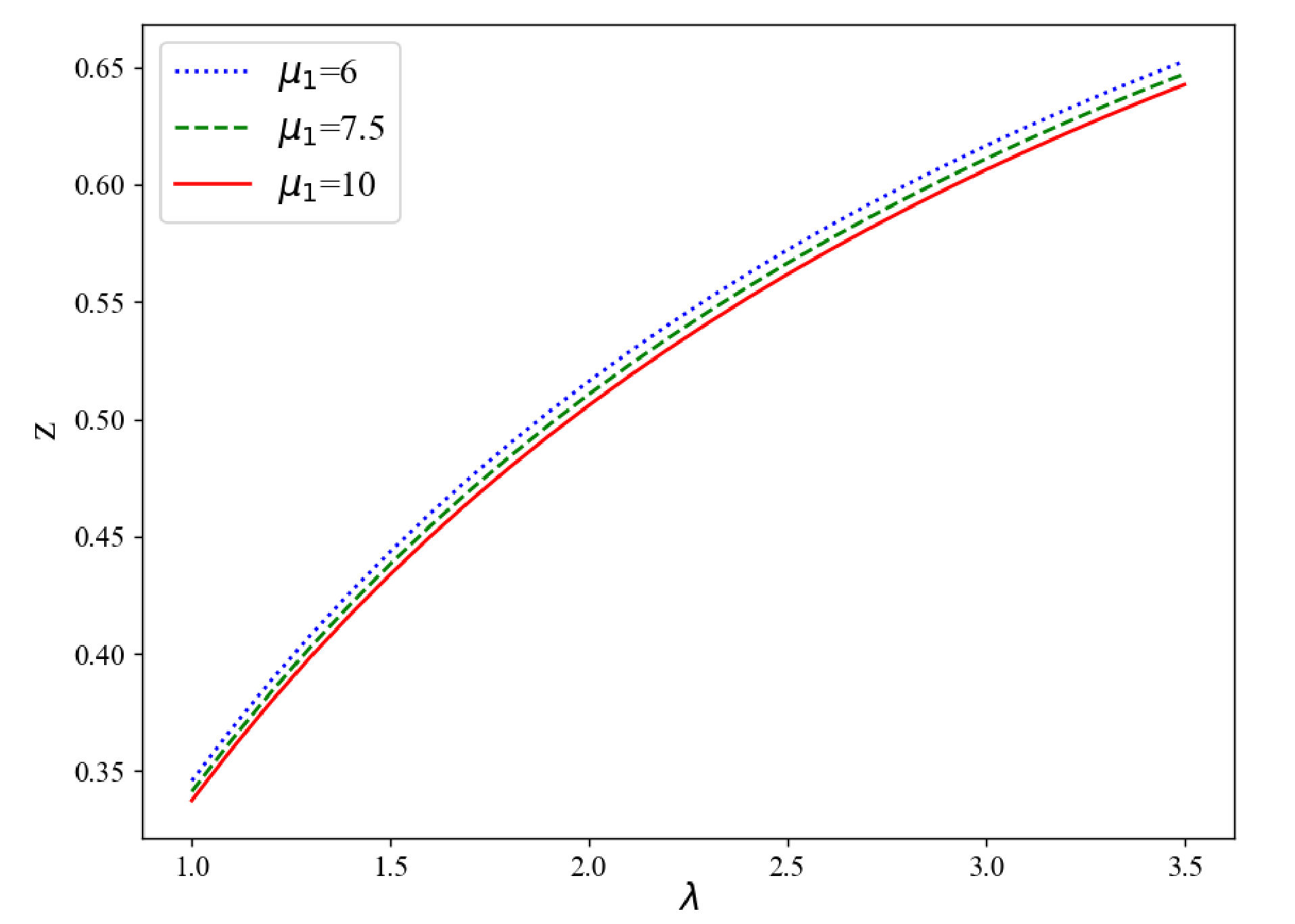}
\caption{$z$ vs. $\lambda$ for three different values of $\mu_{1}$.}%
\end{figure}
\begin{figure}[htbp]
\centering        \includegraphics[width=9cm]{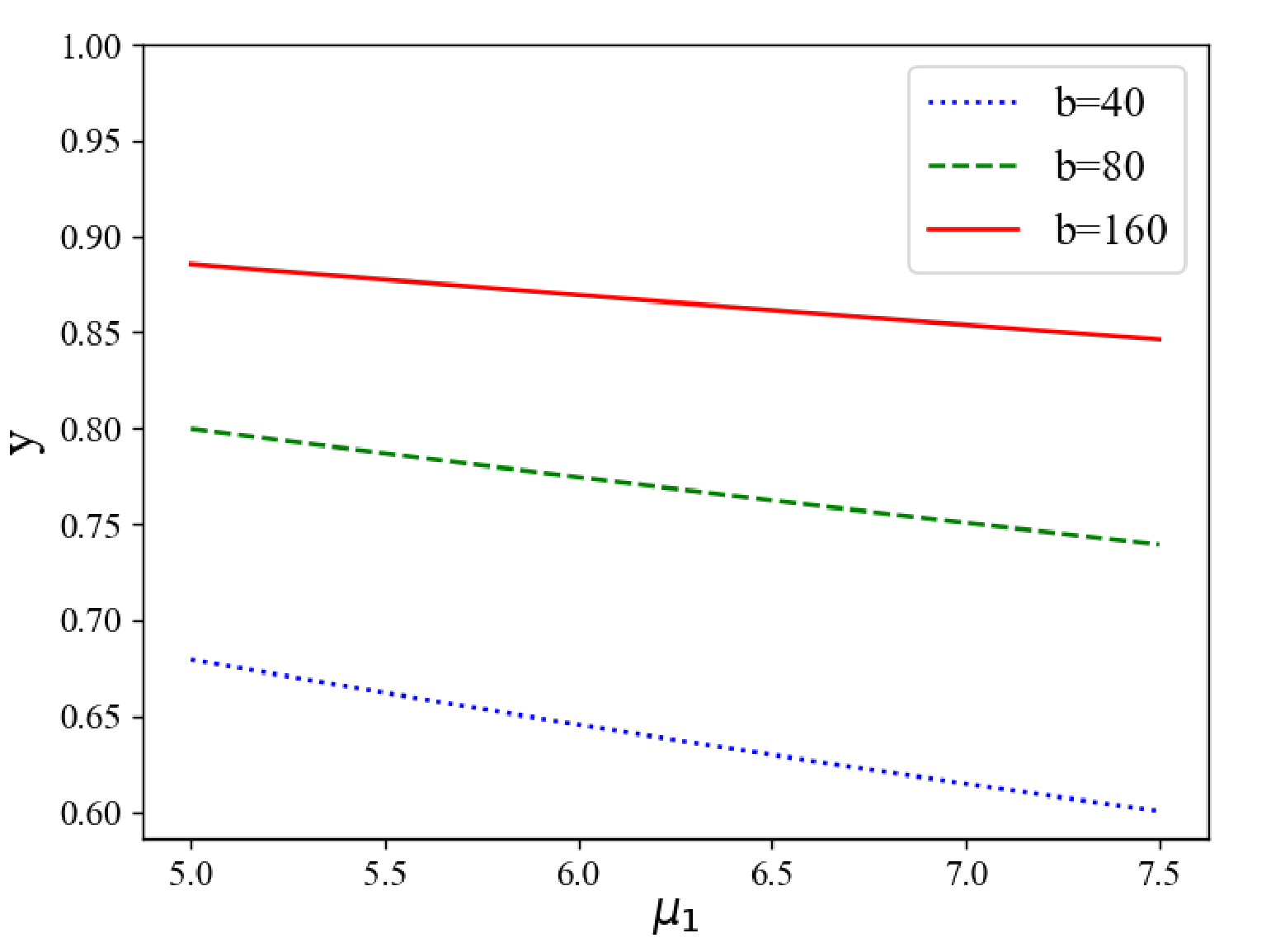}
\caption{$y$ vs. $\mu_{1}$ for three different values of $b$.}
\end{figure}

\textbf{Example 3} We specifically observe how $y$ and $z$ depend on the maximal block size $b$, respectively. We take some common parameters: The arrival rate $\lambda=1.5$, the blockchain-building service rate $\mu_{2}=2$, the maximum block size $b=40, 80, 160$ and the block-generation service rate $\mu_{1}\in\left(  1, 2.5\right)$. From Fig. 4 and Fig. 5, it is seen that $y$ and $z$ decrease as $\mu_{1}$ increases, while they increase as $b$ increases.
\begin{figure}[htbp]
\centering        \includegraphics[width=9cm]{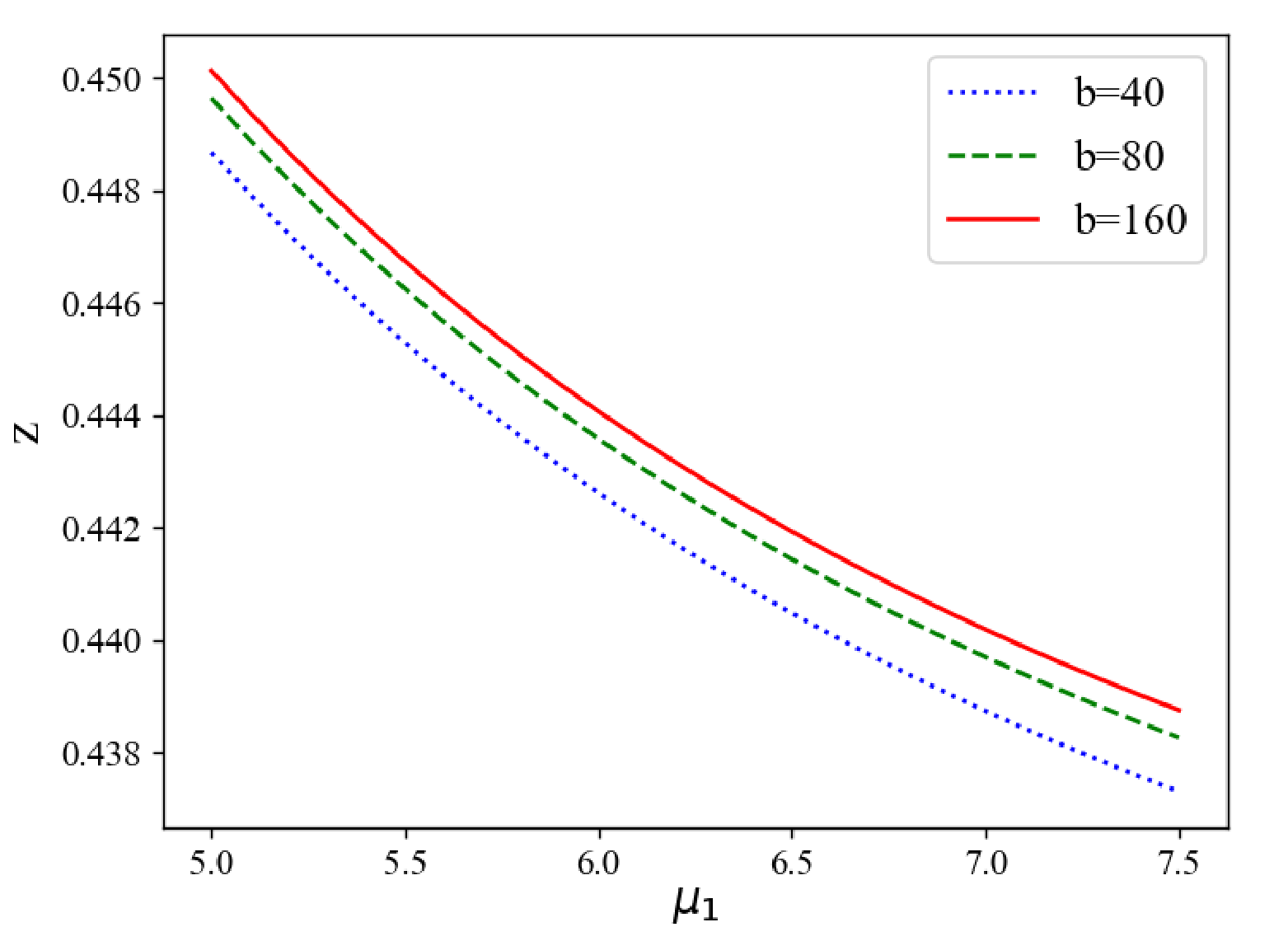}
\caption{$z$ vs. $\mu_{1}$ for three different values of $b$.}
\end{figure}

\section{Concluding Remarks}

In this paper, we apply the maximum entropy principle of the information theory to study the blockchain queueing system, and provide an approximate expression of its steady-state probability distribution. By obtaining this approximation, we have partially resolved a challenging issue in the blockchain technology, i.e., how to directly express the steady-state probability distributions of some large-scale and complex blockchain queueing systems. On the other hand, we use numerical examples to verify the computability of our theoretical results and analyze how the key factors of the approximate expression depend on some crucial parameters. Along these lines, we will continue our future research in the following directions:

$\bullet$ Investigating blockchain queueing systems with multiple mining pools, different consensus mechanisms and so on.

$\bullet$ Extending the information theory to blockchain queueing networks or stochastic models.

$\bullet$ Applying the information theory to provide a more accurate approximate expression with more prior information such as the second moment and the third moment.

\end{document}